\documentclass{pasj00}
\draft

\begin{document}
\SetRunningHead{Omiya et al.}{A Planetary Companion to HD 100655}
\Received{ }
\Accepted{ }

\title{A Planetary Companion to the Intermediate-Mass Giant HD 100655}



%
\author{%
Masashi \textsc{Omiya},\altaffilmark{1,2}
Inwoo \textsc{Han},\altaffilmark{1}
Hideyuki \textsc{Izumiura},\altaffilmark{3,4}
Byeong-Cheol \textsc{Lee},\altaffilmark{1}
Bun'ei \textsc{Sato},\altaffilmark{2}
Kang-Min \textsc{Kim},\altaffilmark{1}
Tae Seog \textsc{Yoon},\altaffilmark{5}
Eiji \textsc{Kambe},\altaffilmark{3}
Michitoshi \textsc{Yoshida},\altaffilmark{6}
Seiji \textsc{Masuda},\altaffilmark{7}
Eri \textsc{Toyota},\altaffilmark{8}
Seitaro \textsc{Urakawa},\altaffilmark{9} and
Masahide \textsc{Takada-Hidai},\altaffilmark{10}}

\altaffiltext{1}{Korea Astronomy and Space Science Institute, 61-1 Whaam-dong, Youseong-gu, Daejeon 305-348, South Korea}
\altaffiltext{2}{Department of Earth and Planetary Sciences, Tokyo Institute of Technology, 2-12-1 Ookayama, Meguro-ku, Tokyo 152-8551, Japan} \email{omiya.m@geo.titech.ac.jp}
\altaffiltext{3}{Okayama Astrophysical Observatory, National Astronomical Observatory of Japan, Asakuchi, Okayama 719-0232, Japan}
\altaffiltext{4}{Department of Astronomical Science, The Graduate University for Advanced Studies, Shonan Village, Hayama, Kanagawa 240-0193, Japan}
\altaffiltext{5}{Department of Astronomy and Atmospheric Sciences, Kyungpook National University, Daegu 702-701, South Korea}
\altaffiltext{6}{Hiroshima Astrophysical Science Center, Hiroshima University, Higashi-Hiroshima, Hiroshima 739-8526, Japan}
\altaffiltext{7}{Tokushima Science Museum, Asutamu Land Tokushima, Itano-gun, Tokushima 779-0111, Japan}
\altaffiltext{8}{Kobe Science Museum, 7-7-6 Minatojimanakamachi, Chuo-ku, Kobe, Hyogo 650-0046, Japan}
\altaffiltext{9}{Bisei Spaceguard Center, Japan Spaceguard Association, 1716-3 Okura, Bisei-cho, Ibara, Okayama 714-1411, Japan}
\altaffiltext{10}{Liberal Arts Education Center, Tokai University, 4-1-1 Kitakaname, Hiratsuka, Kanagawa 259-1292, Japan}

\maketitle

\begin{abstract}
 A precise radial velocity survey conducted by a Korean$-$Japanese planet search program revealed a planetary companion around the intermediate-mass clump giant HD 100655. The radial velocity of the star exhibits a periodic Keplerian variation with a period, semi-amplitude and eccentricity of 157.57 d, 35.2 m s$^{-1}$ and 0.085, respectively. Adopting an estimated stellar mass of 2.4 $M_{\odot}$, we confirmed the presence of a planetary companion with a semi-major axis of 0.76 AU and a minimum mass of 1.7 $M_{\mathrm{J}}$. The planet is the lowest-mass planet yet discovered around clump giants with masses greater than 1.9 $M_{\odot}$. 
\end{abstract}

\section{Introduction}
Over 550 exoplanets have been discovered to date. Many of the planets orbit solar-mass (0.7$-$1.5 $M_{\odot}$) stars, and they have revealed properties that are now used to constrain planet-formation models (e.g., \cite{Ida2004}; \cite{Butler2006}; \cite{Udry2007}). In contrast, only about 60 and 25 planets have been detected around evolved G-K (sub)giants (1.5$-$5 $M_{\odot}$) and K-M dwarfs ($<$0.7 $M_{\odot}$), respectively (e.g., \cite{Sato2008}; \cite{Johnson2011}; \cite{Johnson2007a}). Accordingly, the properties of the planetary systems orbiting such stars are less clarified yet than those for solar-mass stars. Planetary formation depends on the properties of protoplanetary disks, which should be affected by properties of the host star, such as stellar metallicity, radiation output, and disk diffusion times (e.g., \cite{Kornet2006}; \cite{Kennedy2008}). Observational features of planetary systems over a wide range of host star masses need to be clarified by current and future surveys of various masses stars in order to understand planetary formation in general.

More than 20 years ago, initial theoretical ideas of planetary formation for systems over a wide range of stellar masses were presented in terms of planet formation in protoplanetary disks with different properties (\cite{Nakano1988a}; \cite{Nakano1988b}). In the last two decades, improvements in planet formation modeling have made it possible to compare theoretical models directly with observed properties of planetary systems around stars with various masses (e.g., \cite{Ida2005}; \cite{Burkert2007}). For example, \citet{Kennedy2008} predicted that the peak occurrence rate of giant planets occurs for stars with masses of around 3 $M_{\odot}$, based on a core accretion scenario which includes the movements of snow lines under the evolution of central stars. Moreover, \citet{Currie2009} suggested that "the planet desert", i.e., a dearth of planets with semi-major axes of $<$0.6 AU orbiting $>$1.5 $M_{\odot}$ stars, may be reproduced by the effects of Type-II migration, considering the dependence of diffusion time of the protoplanetary disk on stellar mass. Clarifying the relationship between stellar mass and planetary system will provide valuable insights into planet formation models.

For intermediate-mass stars on the main sequence, precise Doppler surveys are difficult because of their large intrinsic radial velocity variations and smooth spectra with few absorption lines, caused by high surface activity, high surface temperature and/or high rotational velocity \citep{Lagrange2009}. In contrast, evolved intermediate-mass (sub)giant stars are suitable targets for precise Doppler surveys because these stars have low surface activity and their spectra exhibit many sharp absorption lines. Thus, to date, spectroscopy-based planet searches targeting intermediate-mass stars have been carried out through precise Doppler surveys of evolved stars. Although the number of substellar companions found orbiting such stars is still insufficient, some characteristic planetary system properties across a wide range of host star masses have begun to emerge. For example, the masses of planets and their host stars show correlation: more massive substellar companions tend to exist around more massive stars (e.g., \cite{Lovis2007}). This correlation suggests that the mass range of the brown dwarf desert depends on host-star's mass, and that planets may be deficient around 2.4$-$4 $M_{\odot}$ stars \citep{Omiya2009}. Also, the planet occurrence rate depends on host-star's mass: the giant planet frequency for higher-mass giant stars is higher than that for lower-mass stars (\cite{Lovis2007}; \cite{Johnson2007a}). The fraction of giant planets increases with increasing stellar mass up to 2 $M_{\odot}$ \citep{Johnson2010a}. Moreover, the orbital semi-major axes of planetary systems also seem to be correlated to host-star's properties. Semi-major axes of most planets orbiting intermediate-mass (sub)giant stars are larger than 0.6 AU\footnote[1]{A planet with a semi-major axis of 0.081 AU was found orbiting an intermediate-mass subgiant star HD 102956 with a mass of 1.68 $M_{\odot}$ \citep{Johnson2010b}.}, while those orbiting solar-type stars are larger than 0.02 AU (\cite{Johnson2007b}; \cite{Sato2008}; \cite{Wright2009}, \cite{Bowler2010}). Even considering the effect of engulfment of inner-orbit planets by host stars, which have experienced rapid expansion in the red giant branch (RGB) phase (\cite{Sato2008}, \cite{Villaver2009}), the observed properties of substellar systems orbiting intermediate-mass (sub)giant stars seem to be different from those orbiting solar-type stars (see also \cite{Bowler2010}).

In 2005, we started a Doppler spectroscopy-based survey of evolved GK-type giants in a framework of a Korean$-$Japanese planet search program \citep{Omiya2009}. The survey program is an extension to the ongoing Okayama Astrophysical Observatory (OAO) planet search program \citep{Sato2005}, and aims to clarify the properties of their associated planetary systems in collaboration with an East-Asian Planet Search Network (EAPS-Net; \cite{Izumiura2005}). About 190 sample stars of the survey were selected from the $Hipparcos$ catalog based on the same criteria as those for OAO planet search program, except visual magnitude (6.2 $<$ $V$ $<$ 6.5). The radial velocity variability of each sample star is monitored using either the 1.8-m telescope at Bohyunsan Optical Astronomy Observatory (BOAO, Korea) or the 1.88-m telescope at OAO (Japan). If a sample star exhibits large variations in radial velocity, follow-up observations of the star are performed using both telescopes. 

In this paper, we report the discovery of a planetary companion orbiting the intermediate-mass giant HD 100655. This is the first planet discovered by this Korean-Japanese planet search program. In section 2, we describe our observations and radial velocity measurements from BOAO and OAO data. The properties of the host star and the radial velocity variability are reported in sections 3 and 4, respectively. We discuss possible causes of the radial velocity variation in section 5. In section 6, we consider the implications of this discovery for the current picture of planetary companions around intermediate-mass giant stars.

\section{Observations and Analyses}
\subsection{BOES Observations and Analysis}
Radial velocity observations at BOAO were carried out with the 1.8-m telescope and the BOAO Echelle Spectrograph (BOES; \cite{Kim2007}), a fiber-fed high resolution echelle spectrograph. We placed an iodine (I$_{2}$) cell in the optical path in front of the fiber entrance of the spectrograph \citep{Kim2002} for precise wavelength calibration and used a 200-$\mu$m fiber, obtaining a wavelength resolution $R$ = $\lambda$/$\Delta{\lambda}$ $\sim$ 51,000. The spectra covered a wavelength region from 3500 $\mathrm{\AA}$ to 10,500 $\mathrm{\AA}$. Echelle data reduction was performed using the IRAF\footnote[2]{IRAF are distributed by the National Optical Astronomy Observatories, which are operated by the Association of Universities for Research in Astronomy, Inc. under cooperative agreement with the National Science Foundation, USA.} software package in the standard manner. We used a wavelength region of 5000$-$5900 $\mathrm{\AA}$ which is covered by many I$_{2}$ absorption lines, for radial velocity measurements. We also made use of Ca II H line at around 3970 $\mathrm{\AA}$ as chromospheric activity diagnostics. Radial velocity analysis was performed using the spectral modeling technique described in \citet{Sato2002}, which was based on the method of \citet{Butler1996} and was adapted to BOES data analysis \citep{Omiya2009}. We employed the extraction method described in \citet{Sato2002} to prepare a stellar template spectrum from stellar spectra taken through the I$_{2}$ cell (star+I$_{2}$ spectra). The technique allowed us to achieve a long-term Doppler precision of 14 m s$^{-1}$ over 4.5 years. 

\subsection{HIDES Observations and Analysis}
Radial velocity observations at OAO were carried out with the 1.88-m telescope and HIgh Dispersion Echelle Spectrograph (HIDES; \cite{Izumiura1999}) attached to the coud\'e focus of the telescope. We used an I$_{2}$ cell placed in the optical path in front of the slit of the spectrograph \citep{Kambe2002} as a precise wavelength calibrator. We always set the slit width to 200 $\mu$m (0.76"), providing a spectral resolution of 63,000. Until November 2007, we had taken star+I$_{2}$ spectra with a wavelength region of 5000$-$6200-$\mathrm{\AA}$. Since the HIDES CCD system was upgraded to a three-CCD mosaic in December 2007, we have obtained spectra from 3750 $\mathrm{\AA}$ to 7550 $\mathrm{\AA}$.  The wavelength region of 5000$-$5900 $\mathrm{\AA}$ of the star+I$_{2}$ spectra are used for radial velocity measurements. The full range of stellar spectra taken without the I$_{2}$ cell are used for abundance analysis. Echelle data reduction was performed using the IRAF software package in the standard manner. Stellar radial velocities were derived from the star+I$_{2}$ spectra using the spectral modeling techniques detailed in \citet{Sato2002}, giving a Doppler precision of less than 8 m s$^{-1}$ over 4.5 years.

\section{Stellar Parameters of HD 100655}
\begin{figure}
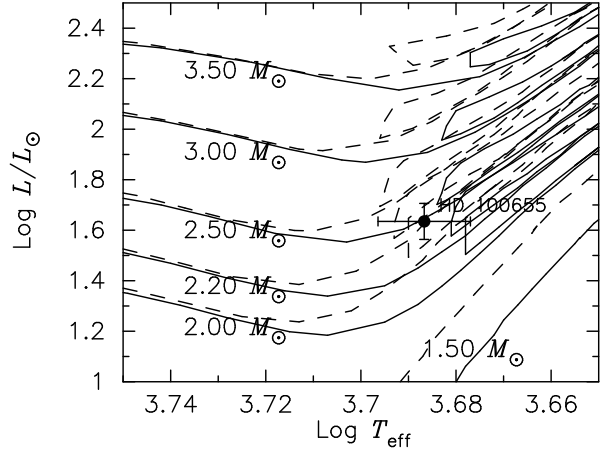

  \begin{center}
    \FigureFile(80mm,60mm){figure1.eps}
  \end{center}
  \caption{HR diagram including HD 100655 with evolutionary tracks \citep{Girardi2000} for Z = 0.03 (solid lines) and Z = 0.019 (dathed lines) for masses of 1.5-3.5 $M_{\odot}$.}
\label{fig1}
\end{figure}

HD 100655 (HR 4459, HIP 56508, BD+21 2331) is 122.3 $\pm$ 7.5 pc from the Sun according to the $Hipparcos$ parallax of $\pi$ = 8.18 $\pm$ 0.50 mas \citep{vanLeeuwen2007}. The star is classified as a G9III giant star with $V$ = 6.45 and $B-V$ = 1.010 $\pm$ 0.015 \citep{ESA1997}. We corrected the observed color index by an extinction value of $E(B-V)$ = 0.0163 $\pm$ 0.0016. The value was calculated from the galactic extinction of $E(B-V)_{S}$ = 0.0273 $\pm$ 0.0015 to the direction of the star obtained from the \citet{Schlegel1998} dust maps using the relation $E(B-V)$ = $E(B-V)_{S}$[1$-$exp($-|D$sin$b|$/125)], where $D$ and $b$ are the distance from sun and the galactic latitude, respectively. We derived an effective temperature of the star of $T_\mathrm{eff}$ = 4861 $\pm$ 110 K using the $(B-V)-T_{\mathrm{eff}}$ calibration of Alonso et al. (\yearcite{Alonso1999}, \yearcite{Alonso2001}). A luminosity of $L$ = 43 $\pm$ 5 $L_{\odot}$ was obtained from the absolute magnitude $M_{V}$ = 0.96 $\pm$ 0.13 and the bolometric correction $B.C.$ = $-$0.31 $\pm$ 0.04 based on the calibration of \citet{Alonso1999}. A stellar mass of $M$ = 2.4$^{+0.2}_{-0.4}$ $M_{\odot}$ was estimated by interpolating the evolutionary tracks of \citet{Girardi2000} with the estimated $T_{\mathrm{eff}}$ and $L$ (see figure \ref{fig1}). We determined the surface gravity to be log $g$ = 2.89 $\pm$ 0.10 and the stellar radius $R$ = 9.3$^{+1.3}_{-1.1}$ $R_{\odot}$ from $M$, $L$, and $T_{\mathrm{eff}}$. The microturbulent velocity $V_{t}$ = 1.36 $\pm$ 0.03 km s$^{-1}$ and the [Fe/H] of 0.15 $\pm$ 0.12 were derived from abundance analysis of a model atmosphere \citep{Kurucz1993} using the equivalent widths of Fe I and Fe II lines measured from an I$_{2}$-free spectrum of HD 100655. We adopted gf-values of Fe I and Fe II lines from \citet{Takeda2005}. \citet{deMedeiros1999} found the stellar rotational velocity, $v\mathrm{sin} i_{s}$, to be 1.6 $\pm$ 1.0 km s$^{-1}$. This value is comparable to the rotational velocities of typical late G-type giants. The stellar parameters are summarized in table \ref{tab1}.

\section{Orbital Solution}
\begin{figure}
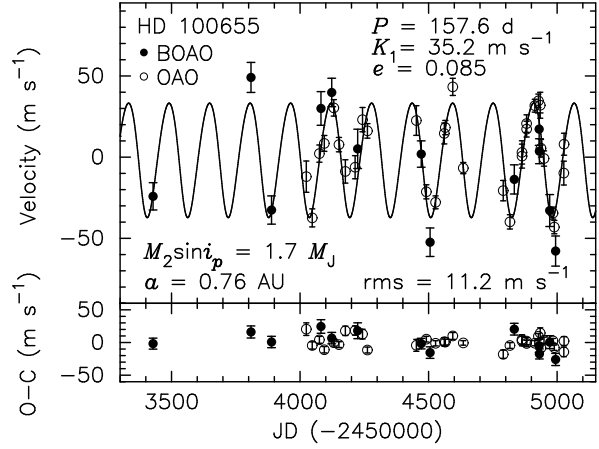

  \begin{center}
    \FigureFile(80mm,60mm){figure2.eps}
  \end{center}
  \caption{Upper panel: radial velocities of HD 100655 observed at BOAO ($filled$ $circles$) and OAO ($open$ $circles$). The solid line represents the Keplerian orbital curve. Lower panel: Residuals to the best Keplerian fit.}
  \label{fig2}
\end{figure}

\begin{figure}
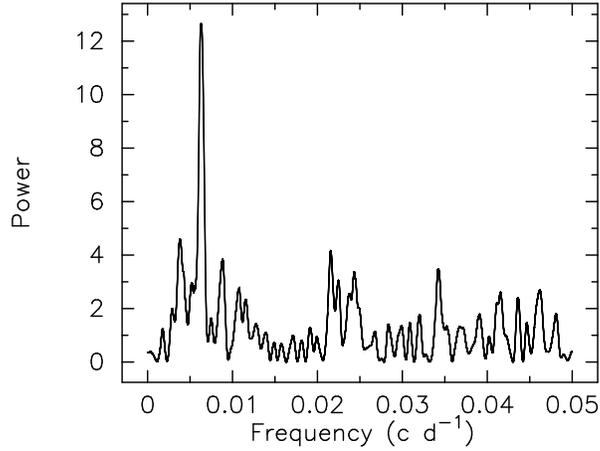

  \begin{center}
    \FigureFile(80mm,60mm){figure3.eps}
  \end{center}
  \caption{The Lomb-Scargle periodgram of the radial velocity variation of HD 100655. A dominant peak appears at a period of 157.78 d (a frequency of 0.006338 c d$^{-1}$) with a False Alarm Probability ($FAP$) of 10$^{-5}$.}
  \label{fig3}
\end{figure}

A large radial velocity variation in the star HD 100655 was found in the early BOAO survey and we made intensive follow-up observations of the star at BOAO and OAO. For 4.5 years from the beginning of the survey, we collected 13 BOAO data points having a typical signal-to-noise ratio (S/N) of 170 pixel$^{-1}$ with an exposure time of 900$-$1200 s, and 32 OAO data points having a typical S/N of 120 pixel$^{-1}$ with an exposure time of 1200$-$1800 s. The observed radial velocities of HD 100655 are shown in figure \ref{fig2} and listed in table \ref{tab2}, together with the observation dates (JD) and estimated uncertainties. A dominant peak in the Lomb-Scargle periodogram \citep{Scargle1982} of the radial velocity variation exists at a period of 157.78 d (a frequency of 0.006338 c d$^{-1}$) (see figure \ref{fig3}). To check the significance of this periodicity, we estimated a False Alarm Probability ($FAP$) using the bootstrap randomization method. We produced 10$^{5}$ fake data sets by randomly mixing the observed radial velocities with a fixed observation date, and applied the Lomb-Scargle periodogram analysis to them. Only one fake data set showed a periodogram power higher than the observed one. Thus, the $FAP$ of the period is 10$^{-5}$. A best-fit Keplerian orbit derived from both the BOAO and OAO velocity data by a least-squares fit has a period $P$ $=$ 157.57 d, a velocity semi-amplitude $K_{1}$ $=$ 35.2 m s$^{-1}$, and an eccentricity $e$ $=$ 0.085. The best-fit curve is shown in figure \ref{fig2} as a solid line overlaid on the observed velocities. We applied an offset of $\Delta$RV = $-$28.1 m s$^{-1}$ to the BOAO velocity data, estimated concurrently with the orbital fit to a Keplerian model. The offset was required because of difference of velocity zero points between BOES and HIDES data originated from using different stellar templates for each data. The rms of the residuals to the best-fit are 14.9 m s$^{-1}$ for BOAO data, 9.2 m s$^{-1}$ for OAO data, and 11.2 m s$^{-1}$ for combined data sets. In the residuals we could not find any significant periodic variation due to additional companions. The best-fit orbital parameters and their uncertainties are listed in table \ref{tab3}. The uncertainties were estimated using a bootstrap Monte Carlo approach by creating 1000 fake data sets. Adopting a stellar mass $M$ = 2.4$^{+0.2}_{-0.4}$ $M_{\odot}$ for HD 100655, we obtained a semi-major axis $a$ = 0.76$^{+0.02}_{-0.04}$ AU and a minimum mass $M_{\mathrm{2}} \mathrm{sin} i_{p}$ = 1.7$^{+0.1}_{-0.2}$ $M_{\mathrm{J}}$ for the planetary companion.

\section{Cause of the Radial Velocity Variation}
To examine causes of the apparent radial velocity variation other than orbital motion, we checked the Ca II H line and the $Hipparcos$ photometric variation, and performed spectral-line shape analyses using a technique described in \citet{Sato2007} as follows. In the analyses, we investigated the cause of the velocity difference between spectra observed at top and bottom velocity phase.

\begin{figure}
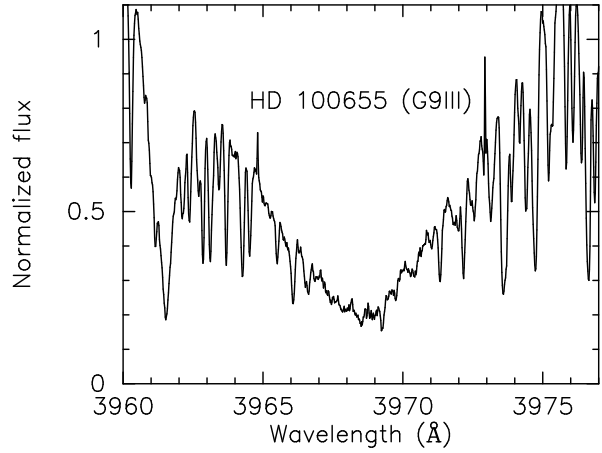

  \begin{center}
    \FigureFile(80mm,60mm){figure4.eps}
  \end{center}
  \caption{The spectrum around the HD 100655 Ca II H line. The line core does not seem to exhibit high chromospheric activity.}
  \label{fig4}
\end{figure}

\begin{figure}
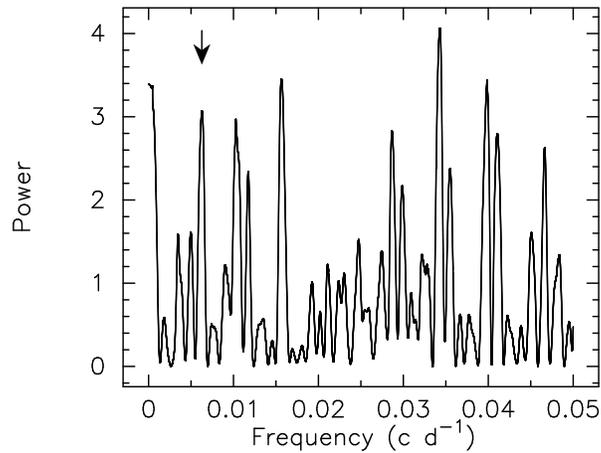

  \begin{center}
    \FigureFile(80mm,60mm){figure5.eps}
  \end{center}
  \caption{Periodgram of $Hipparcos$ photometric variation in HD 100655. Although a weak peak around the period of the stellar radial velocity variation does appear (arrowed line), a $FAP$ of the peak is about 7.4 \%; thus, the peak is not significant one.}
  \label{fig5}
\end{figure}

Figure \ref{fig4} shows the spectrum around the Ca II H line of HD 100655. We note a lack of significant emission in the Ca II H line core of HD 100655, which suggests chromospheric inactivity for the star. Moreover, $Hipparcos$ photometry demonstrates the photometric stability of HD 100655 down to $\sigma$ $\sim$ 0.008 mag. based on the 55 observations for the star over a period of 1000 d. Figure \ref{fig5} displays a periodogram of the $Hipparcos$ photometry. We note a weak peak around the period of the radial velocity variation. To check the significance of the peak, we estimated $FAP$ using the bootstrap method as well as the method described in section 4. We produced 10$^{5}$ fake data sets, and applied the Lomb-Scargle periodogram analysis to them. A total of 7425 fake datasets showed a peak around the period of the radial velocity variation higher than the peak on the observed data set, which means $FAP$ of the peak is about 7.4 \%. Thus the peak is not considered to be significant. Although we have not completely disproved the possibility that the radial velocity variation is due to rotational modulation, these photometric results suggest that the main cause of the observed radial velocity variation is not rotational modulation of stellar spots.

\begin{figure}
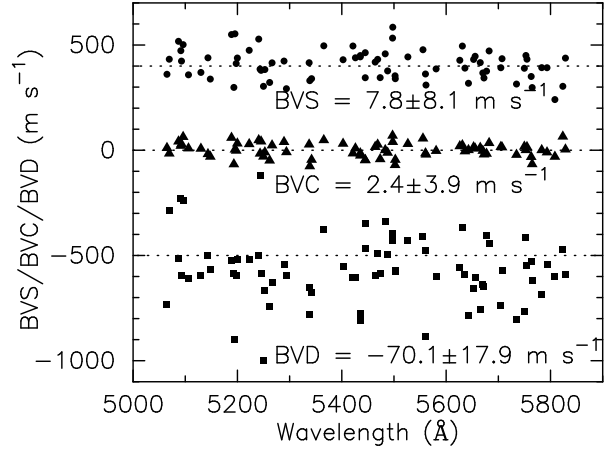

  \begin{center}
    \FigureFile(80mm,60mm){figure6.eps}
  \end{center}
  \caption{Bisector quantities obtained from calculations of cross-correlation functions of two distinct stellar templates. The templates are constructed from star+I$_{2}$ spectra with radial velocities of the peak and valley phase. Values of bisector velocity span (BVS, $circles$), bisector velocity curvature (BVC, $triangles$), and bisector velocity displacement (BVD, $squares$) are shown with offsets of 400 m s$^{-1}$, 0 m s$^{-1}$ and $-$500 m s$^{-1}$, respectively, and their offsets are represented by the dotted-lines.}
  \label{fig6}
\end{figure}

For spectral-line shape analysis, we extracted two high-resolution stellar templates from star(HD 100655)+I$_{2}$ spectra obtained at OAO, using the method described in \citet{Sato2002}. One template was constructed from four spectra with observed radial velocities of the peak phase ($\sim$32 m s$^{-1}$), and the other from four spectra of the valley phase ($-$44 m s$^{-1}$ to $-$34 m s$^{-1}$). Cross-correlation profiles of the two templates were provided for 75 spectral segments (4-\AA \ to 5-\AA \ width each) that did not include severely blended lines or broad lines. We obtained a bisector for the cross-correlation profile of each segment and calculated three quantities from velocities at three flux levels (25\%, 50\%, and 75\%) of the bisector profile. One quantity is the bisector velocity span (BVS), which is the velocity difference between two flux levels with 25\% and 75\% of the bisector. Another is the bisector velocity curvature (BVC), which is the difference between two velocity spans in the upper half (between two flux levels with 50\% and 75\% of the bisector) and the lower half (25\% and 50\%). The other is the bisector velocity displacement (BVD), which is the average of the velocities at the three flux levels (25\%, 50\%, and 75\%). These bisector quantities for HD 100655 are shown in figure \ref{fig6}. The average values of BVS and BVC are 7.8 $\pm$ 8.1 m s$^{-1}$ and 2.4 $\pm$ 3.9 m s$^{-1}$, respectively. The BVS values may be increased due to rotational stellar spots, which may invoke photometric variation. However, since the average value of the BVS is one ninth of the velocity differences ($\sim$70 m s$^{-1}$) between the two templates, we consider both BVS and BVC value to be essentially zero, meaning that the cross-correlation profiles are symmetric. Moreover, the average value of the BVD ($-$70.1 $\pm$ 17.9 m s$^{-1}$) is consistent with the velocity difference between the two templates. Thus, the cause of the velocity difference is considered to be a parallel shift of spectral lines, not variations in spectral line shapes. Hence, the observed radial velocity variation of HD 100655 is best explained by the orbital motion of a planetary companion, not by intrinsic activity, such as rotational modulation and pulsation.

\section{Discussion}
We detected a planetary companion orbiting the clump giant star HD 100655 based on the precise Doppler spectroscopy survey conducted by the Korean$-$Japanese planet search program. The radial velocity variation of the star discovered during early observation at BOAO indicated the existence of a possible planetary companion, and the orbital parameters of the companion were determined by follow-up observations at BOAO and OAO. Adopting a mass of 2.4 $M_{\odot}$ for HD 100655, we found that the planetary companion has a minimum mass of 1.7 $M_{\mathrm{J}}$ and a semi-major axis of 0.76 AU. This is the lowest-mass planet among those discovered around giant stars with masses larger than 1.9 $M_{\odot}$. Fourteen planetary companions and six brown dwarf-mass companions have been detected so far around such giants by ongoing precise Doppler surveys, and the discoveries bring out some characteristic properties of the planetary systems. 

\begin{figure}
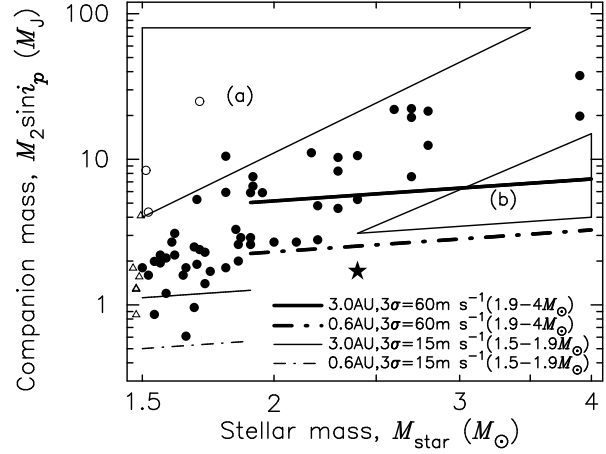

  \begin{center}
    \FigureFile(80mm,60mm){figure7.eps}
  \end{center}
  \caption{Planetary mass and stellar mass of the planetary systems. Solid dots, circles, and triangles represent planetary systems orbiting intermediate-mass (sub)giants, intermediate-mass dwarfs and solar mass stars, respectively. A star indicates the planetary system orbiting HD 100655. Dot-dashed and solid lines mark the detection limits of companions orbiting stars of any mass at 0.6 and 3 AU. Two of the unpopulated regions shown by \citet{Omiya2009} are indicated by (a) and (b). The planetary system of HD100655 appears below the detection limit, because of its small orbital semi-major axis and low radial velocity jitter.}
  \label{fig7}
\end{figure}

Figure \ref{fig7} plots mass of substellar companions with semi-major axes less than 3 AU against host-star's mass (updated version of figure 5 of \cite{Omiya2009}). This figure includes intermediate-mass (1.5 $M_{\odot}$ $\leq$ $M$ $\leq$ 4 $M_{\odot}$) giants and subgiants ($filled$ $circles$), intermediate-mass dwarfs ($open$ $circles$), solar-mass stars ($M$ $<$ 1.5 $M_{\odot}$, $open$ $triangles$), and HD 100655 ($star$). Solid and dot-dashed lines indicate the lower-mass limits of companions detectable by current Doppler surveys for semi-major axes of 0.6 AU and 3 AU, respectively. The limits correspond to companion masses that give rise to semi-amplitudes of radial velocity variations of their host stars as large as three times the typical radial velocity jitters, which are 5 m s$^{-1}$ for subgiants (1.5$-$1.9 $M_{\odot}$) and 20 m s$^{-1}$ for clump giants (1.9$-$4 $M_{\odot}$) (\cite{Johnson2010c}, \cite{Sato2005}). Two unpopulated regions of substellar companions orbiting intermediate-mass subgiants and giants appear in regions (a) and (b) in figure \ref{fig7} \citep{Omiya2009}. The planet orbiting HD 100655 is located below the detection limit and the region (b) in figure \ref{fig7} because of its small root mean-square scatter of the residual radial velocities ($\sim$11 m s$^{-1}$), that is, its small radial velocity jitter. The existence of this planet suggests a possibility that low-mass giant planets can form around $\sim$2.4 $M_{\odot}$ stars, while this planet could possibly have a small orbital inclination, and thus a high actual mass. Therefore, a paucity of low-mass companions orbiting massive intermediate-mass giants, roughly indicated by the region (b), might partly be caused by an observational bias due to the high detection limit. In this respect, observational surveys more sensitive to lower-mass substellar companions are necessary.

\begin{figure}
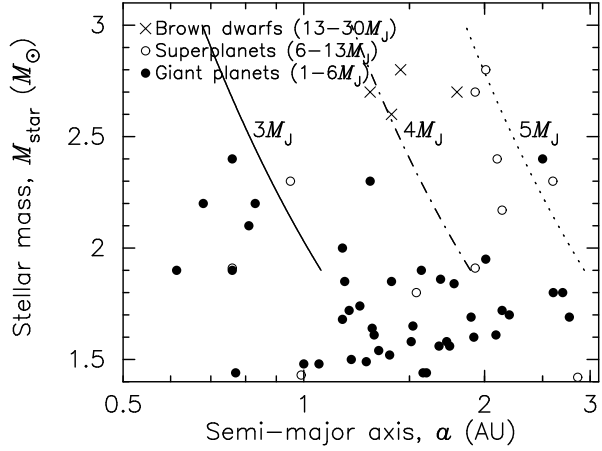

  \begin{center}
    \FigureFile(80mm,60mm){figure8.eps}
  \end{center}
  \caption{Orbital semi-major axis versus stellar mass of planetary systems. Dots, circles, and crosses indicate the locations of giant planets (1$-$6 $M_{\mathrm{J}}$), superplanets (6$-$13 $M_{\mathrm{J}}$), and brown dwarfs (13$-$30 $M_{\mathrm{J}}$), respectively. Solid, dot-dashed and dotted lines indicate the typical largest semi-major axes of companions detectable by current Doppler surveys with masses of 3, 4 and 5 $M_{\mathrm{J}}$, respectively. Many planets around 1.9$-$2.5 $M_{\odot}$ giant stars seem to belong to normal giant planets orbiting at semi-major axes of 0.6$-$1.3 AU, or superplanets orbiting at semi-major axes of 1.9$-$3 AU.}
  \label{fig8}
\end{figure}

The mass distribution of substellar companions orbiting 1.5$-$3 $M_{\odot}$ stars may also depend on the semi-major axes of the companions. Figure \ref{fig8} is a plot of semi-major axis of substellar companion versus host star mass. Crosses, circles and filled circles indicate brown dwarf-mass companions (13$-$30 $M_{\mathrm{J}}$), "superplanets" (6$-$13 $M_{\mathrm{J}}$), and normal giant planets (1$-$6 $M_{\mathrm{J}}$), respectively. Solid, dot-dashed and dotted lines indicate the typical farthest orbital distances of companions detectable by current Doppler surveys for companion masses of 3, 4 and 5 $M_{\mathrm{J}}$, respectively. The distances correspond to orbital semi-major axes that the companions induce radial velocity variations of their host stars with semi-amplitudes as large as three times the typical radial velocity jitter, which is 20 m s$^{-1}$ for clump giants (1.9$-$3 $M_{\odot}$) \citep{Sato2005}. In figure \ref{fig8}, some interesting properties of substellar companions are suggested in three stellar mass ranges. Almost all the planets orbiting 1.5$-$1.9 $M_{\odot}$ stars are normal giant planets, and are located on orbits with semi-major axes of $>$1 AU. Many planets orbiting 1.9$-$2.5 $M_{\odot}$ stars seems to be classified in two groups\footnote[3]{We note that $<$3 $M_{\mathrm{J}}$ ($<$5 $M_{\mathrm{J}}$) planets orbiting such stars at 1 AU (3 AU) are below the lower-mass limits for typical detectable planets.}: normal giant planets at inner orbits (0.6$-$1.3 AU) and superplanets at outer orbits (1.9$-$3 AU). HD 100655 b is included in the group of the normal giant planets. All planet-mass companions orbiting 2.5$-$3 $M_{\odot}$ stars reside at semi-major axes larger than 1.9 AU, while all brown dwarf-mass companions are orbiting at semi-major axes less than 1.9 AU. Although the number of known substellar companions discovered around stars with $>$2.5 $M_{\odot}$ is still small, the distribution of substellar companions around 1.9$-$2.5 $M_{\odot}$ stars may differ from those around 1.5$-$1.9 $M_{\odot}$ and 2.5$-$3 $M_{\odot}$ stars.

To reproduce the distribution of giant planets around 1.9$-$2.5 $M_{\odot}$ giant stars, two scenarios can be suggested. One is the planet engulfment scenario caused by stellar evolution of primary stars. Most of the host stars are clump giants that should have experienced the RGB phase, which triggers rapid stellar expansion. \citet{Villaver2009} suggested that the primary stars can preferentially capture more massive planetary companions by tidal interaction in the RGB phase. Thus, the superplanets with semi-major axes of $<$1.9 AU might have been preferentially engulfed by their primary stars even if they had existed, leaving normal giant planets. However, according to \citet{Kunitomo2011}, the critical semi-major axis, within which a primary star can engulf planetary companions, decreases from $\sim$1.5(0.4) AU for 1.7(2.0) $M_{\odot}$ stars to $\sim$0.2 AU for 2.1 $M_{\odot}$ stars and thus all the planets with semi-major axis larger than 0.6 AU around 2.0$-$2.5 $M_{\odot}$ stars can survive the RGB phase regardless of their masses. Therefore, the observational properties would not be quantitatively explained by only this mechanism. 

The other scenario is that the distribution of the planets is primordially originated from planet migration in protoplanetary disk. Dependence of Type-II migration rate on planet mass may separate locations of low-mass giant planets and superplanets. For example, based on the equation (1) of \citet{Currie2009}, 2 $M_{\mathrm{J}}$ and 8 $M_{\mathrm{J}}$ planets that formed in circular orbits with a semi-major axis of 3.8 AU around 2 $M_{\odot}$ stars can migrate to inner orbits with semi-major axes of $\sim$0.7 AU and $\sim$2.8 AU, respectively, assuming a disk dissipation time of 1 Myr. Thus, the observed orbital distribution of planets around 1.9$-$2.5 $M_{\odot}$ stars may be explained by this mechanism, and if this is the case many undetected lower-mass planets should be expected at distances of 1$-$3 AU because giant planets can form at any distance beyond the snow line. It should be noticed, however, the observed semi-major axis distributions of planetary systems around 1.5$-$1.9 $M_{\odot}$ and 2.5$-$3 $M_{\odot}$ stars might not be explained by only the effect of the migration. 

Additionally, Type-II migration may not be only the mechanism that can locate giant planets around intermediate-mass stars. The magnetorotational instability-dead zone in the protoplanetary disks may encourage formations of giant planets only at $\sim$1 AU around intermediate-mass stars \citep{Kretke2009}. In this case, a drop-off of giant planets at $>\sim$1 AU might exist around such stars. However, the distribution of $<$3 $M_{\mathrm{J}}$ planets at larger than 1 AU around 1.9$-$3 $M_{\odot}$ stars has not been clarified yet due to the detection limits of current planet searches. 

Thus, in order to examine roles of these mechanisms on planet formation and evolution around intermediate-mass stars, it is required to evaluate the semi-major axes distribution by further Doppler surveys of intermediate-mass stars with masses of $>$1.9 $M_{\odot}$ sensitive to lower-mass planets.

\bigskip
This research was supported as a Korea-Japan Joint Research Project under the Japan-Korea Basic Scientific Cooperation Program between Korea Science and Engineering Foundation (KOSEF) and Japan Society for the Promotion of Science (JSPS). This research is based on data collected at Bohyunsan Optical Astronomy Observatory (BOAO) that is operated by Korea Astronomy and Space Science Institute (KASI) and Okayama Astrophysical Observatory (OAO) that is operated by National Astronomical Observatory of Japan (NAOJ). We gratefully acknowledge the support from the staff members of BOAO and OAO during the observations. BCL acknowledges the support from Engineering Foundation (KOSEF) through the Science Research Center (SRC) program. TSY acknowledges the support from Basic Science Research Program through the National Research Foundation of Korea (NRF) funded by the Ministry of Education, Science and Technology (No. 2010-0023430). Data analyses were in part carried out on common use data analysis computer system at the Astronomy Data Center, ADC, of the National Astronomical Observatory of Japan. This research has made use of the SIMBAD database, operated at CDS, Stransbourg, France.


\bigskip

\begin{table}
  \caption{Stellar parameters of HD 100655}\label{tab1}
  \begin{center}
    \begin{tabular}{ll}
      \hline
      Parameter &   Value \\
      \hline
Spectral Type & G9III \\
$V$ & 6.45 \\
$B-V$ & 1.010 $\pm$ 0.015 \\
$\pi$ (mas) & 8.18 $\pm$ 0.50 \\
$M_{V}$ & 0.96 $\pm$ 0.13 \\
$B.C.$ & -0.31 $\pm$ 0.04 \\
$T_{\mathrm{eff}}$ (K) & 4861 $\pm$ 110 \\
$L$ ($\LO$) & 43 $\pm$ 5 \\
$M$ ($\MO$) & 2.4$^{+0.2}_{-0.4}$ \\
$R$ ($\RO$) & 9.3$^{+1.3}_{-1.1}$ \\
log $g$ & 2.89 $\pm$ 0.10 \\
$V_{t}$ (km s$^{-1}$) & 1.36 $\pm$ 0.03 \\
$[$Fe/H$]$ & 0.15 $\pm$ 0.12 \\
$v \mathrm{sin} i_{s}$ (km s$^{-1}$) &  1.6 $\pm$ 1.0\\
      \hline
    \end{tabular}
  \end{center}
\end{table}

\renewcommand{\baselinestretch}{1.1}

\begin{table}
  \begin{center}
    \caption{Radial velocities of HD 100655} \label{tab2}
    \begin{tabular}{lccc}
      \hline
      \multicolumn{1}{c}{JD} & Radial Velocity & Uncertainties & \\
      $-$2450000 & (m s$^{-1}$) & (m s$^{-1}$) & Observatory\\
      \hline
3428.2166 & 3.9 & 8.4 & BOAO \\ 
3809.1443 & 77.2 & 9.2 & BOAO \\ 
3889.0614 & $-$4.5 & 8.7 & BOAO \\ 
4024.3429 & $-$12.0 & 9.6 & OAO \\ 
4047.3387 & $-$37.4 & 5.6 & OAO \\ 
4075.2415 & 2.2 & 5.3 & OAO \\ 
4081.3382 & 58.1 & 10.4 & BOAO \\ 
4094.3171 & 8.5 & 4.9 & OAO \\ 
4123.2074 & 67.9 & 8.8 & BOAO \\ 
4131.2972 & 30.4 & 5.1 & OAO \\ 
4151.2206 & 7.7 & 4.4 & OAO \\ 
4176.2617 & $-$8.8 & 7.2 & OAO \\ 
4214.1201 & $-$6.2 & 7.1 & OAO \\ 
4224.0873 & 33.2 & 12.1 & BOAO \\ 
4243.0406 & 23.1 & 7.5 & OAO \\ 
4262.0463 & 16.2 & 4.5 & OAO \\ 
4452.3084 & 22.6 & 9.1 & OAO \\ 
4471.3229 & 30.0 & 8.1 & BOAO \\ 
4491.2702 & $-$21.4 & 4.3 & OAO \\ 
4505.7574 & $-$24.3 & 8.8 & BOAO \\ 
4527.2340 & $-$27.8 & 4.0 & OAO \\ 
4561.1034 & 14.6 & 6.4 & OAO \\ 
4565.0837 & 18.2 & 4.4 & OAO \\ 
4594.0472 & 43.3 & 5.3 & OAO \\ 
4634.9712 & $-$6.7 & 3.4 & OAO \\ 
4790.3591 & $-$20.6 & 6.3 & OAO \\ 
4816.3500 & $-$39.9 & 4.4 & OAO \\ 
4833.3367 & 14.4 & 9.0 & BOAO \\ 
4863.2825 & 2.9 & 7.2 & OAO \\ 
4863.3410 & 0.6 & 7.4 & OAO \\ 
4881.1750 & 20.7 & 5.4 & OAO \\ 
4881.3355 & 17.3 & 4.9 & OAO \\ 
4913.2182 & 31.7 & 4.0 & OAO \\ 
4927.1213 & 34.7 & 4.2 & OAO \\ 
4930.0152 & 45.3 & 10.0 & BOAO \\ 
4931.0597 & 31.7 & 7.7 & BOAO \\ 
4934.2081 & 32.0 & 8.0 & OAO \\ 
4937.0928 & 5.5 & 5.4 & OAO \\ 
4948.9873 & $-$0.8 & 4.7 & OAO \\ 
4971.0609 & $-$4.8 & 9.8 & BOAO \\ 
4984.0086 & $-$34.5 & 3.7 & OAO \\ 
4988.9737 & $-$43.1 & 4.1 & OAO \\ 
4994.0597 & $-$29.7 & 9.3 & BOAO \\ 
5025.9988 & $-$9.8 & 7.3 & OAO \\ 
5026.9737 & 8.0 & 6.8 & OAO \\      \hline
    \end{tabular}
  \end{center}
\end{table}

\renewcommand{\baselinestretch}{1.5}

\begin{table}
  \caption{Orbital parameters of HD 100655 b}\label{tab3}
  \begin{center}
    \begin{tabular}{ll}
      \hline
      \multicolumn{1}{c}{Parameter} & Value\\
      \hline
$K_{1}$ (m s$^{-1}$) & 35.2 $\pm$ 2.3 \\
$P$ (days) & 157.57 $\pm$ 0.65 \\
$e$ & 0.085 $\pm$ 0.054 \\
$\omega$ (deg) & 132 $\pm$ 37 \\
$T$ (JD) & 2453072.4 $\pm$ 15.9 \\
$\Delta$RV\footnotemark[$*$] (m s$^{-1}$) & $-$28.1 \\
rms (m s$^{-1}$) & 11.2 \\
Reduced $\sqrt{\chi^{2}}$ & 1.6 \\
$N_{obs}$ & 45 \\
$a_{1} \mathrm{sin} i_{p}$ (10$^{-3}$AU) & 0.508$^{+0.034}_{-0.039}$ \\
$f_{1}$($m$) (10$^{-7}M_{\odot}$) & 0.0071$^{+0.0014}_{-0.0014}$ \\
$M_{\mathrm{2}} \mathrm{sin} i_{p}$ ($M_{\mathrm{J}}$) & 1.7$^{+0.1}_{-0.2}$ \\
$a$ (AU) & 0.76$^{+0.02}_{-0.04}$  \\
       \hline
\multicolumn{1}{@{}l@{}}{\hbox to 0pt{\parbox{85mm}{\footnotesize
\footnotemark[$*$]Offset between OAO and BOAO velocities.
}\hss}}
    \end{tabular}
  \end{center}
\end{table}

\end{document}